# Status of MagAO and review of astronomical science with visible light adaptive optics

Laird M. Close[1a], Jared R. Males[a], Katie M. Morzinski[a], Simone Esposito[b],
Armando Riccardi[b], Runa Briguglio[b], Kate B. Follette[c], Ya-Lin Wu[a], Enrico Pinna[b], Alfio Puglisi[b], and Marco Xompero[b], Fernando Quiros[d], Phil M. Hinz[a]

[a] Steward Observatory, University of Arizona, Tucson AZ 85721, USA;
[b] INAF-Osservatorio Astrofisico di Arcetri, Largo E. Fermi 5, I-50125 Firenze, Italy;
[c] Amherst College; [d]GMT Observatory

## ABSTRACT

We review astronomical results in the visible ($\lambda<1\mu m$) with adaptive optics and note the status the MagAO system and the recent upgrade to visible camera's Simultaneous/Spectra Differential Imager (SDI+) mode. Since mid-2013 there has been a rapid increase visible AO with over 50 refereed science papers published in just 2015-2016 timeframe. Currently there are productive small (D < 2 m) visible light AO telescopes like the UV-LGS Robo-AO system (Baranec, et al. 2016). The largest (D=8m) telescope to achieve regular visible AO science is SPHERE/ZIMPOL. ZIMPOL is a polarimeter fed by the ~1.2 kHz SPHERE ExAO system (Fusco et al. 2016). ZIMPOL's ability to differentiate scattered polarized light from starlight allows the sensitive detection of circumstellar disks, stellar surfaces, and envelopes of evolved AGB stars. The main focus of this paper is another large (D=6.5m Magellan telescope) AO system (MagAO) which has been very productive in the visible as well (particularly at the H−alpha emission line). MagAO is an advanced Adaptive Secondary Mirror (ASM) AO system at the Magellan in Chile. This ASM secondary has 585 actuators with < 1 msec response times (0.7 ms typically). MagAO utilizes a 1 kHz pyramid wavefront sensor (PWFS). The relatively small actuator pitch (~22 cm/subap, 300 modes, upgraded to 30 pix dia. PWFS) allows moderate Strehls to be obtained in the visible (0.63-1.05 microns). Long exposures (60s) achieve <30mas resolutions and 30% Strehls at 0.62 microns (r') with the VisAO camera (0.5-1.0 μm) in 0.5" seeing with bright R ≤ 9 mag stars (~10% Strehls can be obtained on fainter R~12 mag guide stars). Differential Spectral Imaging (SDI) at H-alpha has been very important for accreting exoplanet detection. There is also a 1-5micron science camera (Clio; Morzinski et al. 2016). These capabilities have led to over 35 MagAO refereed science publications. Here we review the key steps to having good performance in the visible and review the exciting new AO visible science opportunities and science results in the fields of: exoplanet detection; circumstellar and protoplanetary disks; young stars; AGB stars; emission line jets; and stellar surfaces. The recent rapid increase in the scientific publications and power of visible AO is due to the maturity of the next-generation of AO systems and our new ability probe circumstellar regions with very high (10-30 mas) spatial resolutions that would otherwise require much larger (>10m) diameter telescopes in the infrared.

## 1.0 INTRODUCTION

### 1.1 Review of AO Astronomy in the Visible

In the early years of AO (1990-1994) AO work (often by the military) was done in the visible (Rigaut 2015). The 1.5m Starfire AO telescope started with visible scoring cameras. Strehls were quite low ~10% at 0.85μm, and resolutions were also fairly low at ~0.21-0.25" FWHM (Drummond et al. 1994). The overall use of AO for astronomical science (at any wavelength) was still fairly rare. Moreover, once HST was repaired in late 1993 visible AO observations lost favor with astronomers who switched to using HST. However, the newly available NIR (NICMOS/InSb) arrays for ground-based AO astronomical observations made NIR AO on 4-10m telescopes very complimentary to HST visible resolutions. In this manner there was no direct competition from HST's superior performance in the visible, and AO thrived in the NIR in the late 90's onwards (see Close 2000 for a review).

---

[1] lclose@as.arizona.edu; phone +1 520 626 5992

In the later years (1995-2012) the 1-2.5μm wavelength range (the NIR) proved a great scientific home for AO astronomical science (still mostly independent of direct HST competition) and produced Strehls that were high (~60% at K) on 4-8m class telescopes. However, visible AO had little significant astronomical science impact in "night-time" astronomy (<1 paper/yr). But visible AO was extensively used for Solar, Military SSA at, for example, AEOS with VisIM in Maui. Also the visible has always been home to amateur AO astronomy (often with some form of Lucky imaging and post-detection frame selection; again not the topic of this paper). Nevertheless, there were almost no astronomical refereed papers in the visible from 1994-2012. The most likely reasons for this were: 1) the Strehls were not very good in the visible (often <1%) with only ~100 corrected modes (at just a few hundred Hz WFS frame rate) from the systems of this generation; and 2) all the visible photons were typically used for the wavefront sensor. Visible AO science really required that an AO system be designed from the start to work in the visible (e.g. with <135 nm rms wavefront error – effectively requiring ≥300 of very well corrected modes and at least ~1 kHz WFS framerates for D≥6m telescopes).

**1.2 Scientific Advantages to Visible AO**

Despite its demanding nature, visible AO has many scientific advantages over the NIR. After all, most astronomy is done in the visible, but almost no AO science was done with λ<1μm on large 6.5-10m class telescopes until recently. A short list of some of the advantages of AO science in the visible compared to the NIR are:

-- **Better science detectors** (CCDs): much lower dark current, lower readnoise (<1e- with EMCCDs), much better cosmetics (no bad pixels), ~40x more linear, and camera optics can be warm, simple, and compact.

-- **Much Darker skies:** the visible sky is 100-10,000x darker than the K-band sky.

-- **Strong Emission lines**: access to the primary visible recombination lines of Hydrogen (Hα 0.6563 μm) --- most of the strongest emission lines are all in the visible, and these have the best calibrated sets of astronomical diagnostics. For example, the brightest line in the NIR is Paβ some *20 times less strong* than Hα (in typical Case-B recombination conditions, T~10,000K).

-- **Off the Rayleigh-Jeans tail**: Stars have much greater range of colors in the visible (wider range color magnitudes) compared the NIR which is on the Rayleigh-Jeans tail. Moreover, visible photometry combined with the IR enables extinction and spectral types to be much better estimated.

-- **Higher spatial resolution**: The 20 mas resolution regime opens up. A visible AO system at r band (λ=0.62μm) on a 6.5m telescope has the spatial resolution of ~20 mas (with full *uv* plane coverage unlike an interferometer) that would otherwise require a 24m ELT (like the Giant Magellan Telescope) in the K-band. So visible AO can produce ELT like NIR resolutions on today's 6.5-8m class telescopes.

**1.3 Keys to good AO Performance in the Visible: Some point design considerations**

While it is certainly clear that there are great advantages to doing AO science in the visible it is also true that there are real challenges to getting visible AO to produce even moderate Strehls on large telescopes. The biggest issue is that $r_o$ is small ~15-20 cm in the visible (since $r_o=15(\lambda/0.55)^{6/5}$ cm on 0.75" seeing site). Below we outline (in rough order of importance) the most basic requirements to have a scientifically productive visible AO system on a 6.5m sized telescope:

1. **Good 0.6" V-band Seeing Site** – Large $r_o$ (>15cm at 0.55μm) and consistency (like clear weather) is critical. In particular, low wind (long $\tau_o$ > 5ms) sites are hard to find, and so this drives loop WFS update frequencies ≥1 KHz (due in large part to the fast jet stream level winds).

2. **Good DM and fast non-aliasing WFS**: need many (>500) actuators (with $d<r_{o/2}$ sampling), no "bad/stuck" actuators, need at least ≥300 well corrected modes for D>6m. Currently the pyramid WFS (PWFS) is the best for NGS science (uses the full pupil's diffraction for wavefront error measurement), hence a PWFS is preferred.

3. **Minimize all non-common path (NCP) Errors:** Stiff "piggyback" design with simple visible science camera well coupled to the WFS –keep complex optics (like the ADC) on the common path. Keep optical design simple and as common as possible. Limit NCP errors to less than 30 nm rms. If the NCP errors are >30 nm rms employ an extra non-common path DM to fix these errors feed by a LOWFS sensor (see Guyon et al. 2018; these proc. and Males et al. 2018; these proc.).

4. **Minimize the Low Wind Effect (LWE):** This is a particularly limiting wavefront error that is linked to the strong radiative cooling of the secondary support spiders. It can be mostly removed by adding a low emissivity coating so that the spiders track the temperature of the night air and do not super-cool. Also a pyramid wavefront sensor seems much better at sensing the LWE errors compared to SH WFS (see Milli et al. 2018 this meeting).

5. **Minimize the Isolated Island Effect**: Unfortunately, the push towards having more than 500 corrected modes and having thick stable spider arms forces visible AO systems to subap sizes that approach spiders arm thicknesses. Hence some sub-apertures are completely in the shadow of the spider arms and so cannot be effectively used by the WFS, allowing the DM to "run-away" in piston w.r.t. each quadrant/section of the pupil (see Obereder et al. 2018 these proc.). This is an insidious problem that demands the use of PWFSs (which can sense the phase difference between the quadrants in these dark zones) and perhaps an additional real-time interferometer like a Zernike sensor (ZELDA; N'Diaye et al. 2017; or phase diversity in science focal plane; N'Diaye et al. 2018; these proc.). To date this is not a completely solved problem on-sky, but it will *dominate* the error budget of ELT AO (even in the NIR) without excellent sensing and correction (Obereder et al. 2018).

4. **Lab Testing:** Lots (and lots) of "end-to-end" closed loop testing with visible science camera. Alignment must be excellent and very stiff for all non-common path optics (for all observing conditions) to minimize NCP errors.

5. **Modeling/Design:** Well understood error budget feeding into analytical models, must at least expect ~135 nm rms WFE on-sky. Try to measure/eliminate vibrations from the telescope and environment with advanced rejection/filter techniques (eg. linear quadratic estimation (LQG) filters).

6 **High Quality Interaction Matrixes:** Excellent on-telescope IMATs with final/on-sky pupil. Take IMATs in partial low-order closed-loop to increase the SNR of the high order modes in the IMAT.

7 **IR camera simultaneous with Visible AO camera:** this is important since you achieve a 200% efficiency boost. Allows for excellent contingency in poor seeing when only NIR science is possible.

8. **Leverage Differential Techniques for Enhanced Contrasts:** Since the visible Strehl (~30%) is not high for any current AO systems, contrast gain from coronagraphy is very limited. However, differential techniques such as Spectra Differential Imaging (SDI) or Polarmetric Differential Imaging (PDI) are very effective in the visible, and when combined with Angular Differential Imaging (ADI) observations with Principal Component Analysis (PCA) data reduction techniques, can lead to very high contrast detections of $10^5$ within 0.5" (Males et al. 2014).

## 1.4 Visible AO vs. HST, Interferometers, or NIR AO on 8-10m Telescopes

The Magellan AO system (MagAO) is a good example of a visible AO system. MagAO can regularly obtain (>50% of the nights so far observed) moderate Strehl (~20% at 0.65μm) and 20-30 mas resolution images in the visible (Close et al. 2013; Fig. 1). *This is >2x higher resolution than the Hubble Space Telescope can achieve at the same wavelengths, and is also ~2-3x better than the sharpest images one can make from the ground with conventional NIR AO on the largest 8-10m apertures.* While interferometers can provide higher spatial resolutions, their limited "*uv*" coverage, limiting magnitudes, and very small FOV (<0.1") make them generally much less attractive than direct imaging for the science cases outlined in section 2.3. Also, speckle interferometry can achieve the diffraction-limit, but is only effective on the brightest binary stars in the optical, with limited contrast and, hence, dynamic range. Whereas, MagAO has detected exoplanets $10^{-5}$ times fainter within 0.5" of the host star (Males et al. 2014) such contrasts are impossible to achieve with any speckle or interferometric techniques.

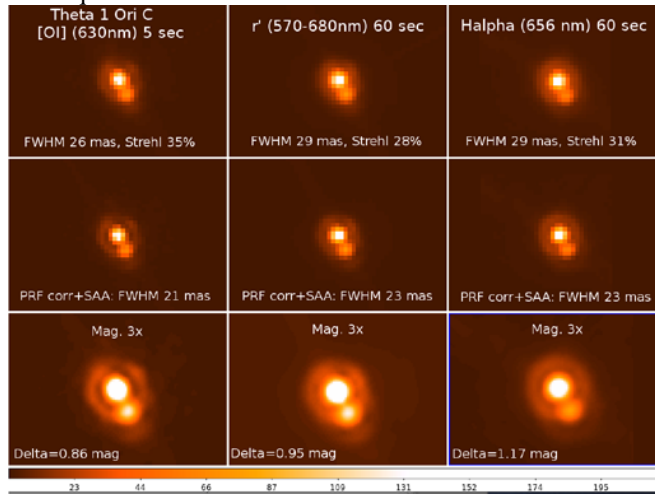

*Fig 1: Top row*: the central ionizing binary of the Trapezium: $\theta^1$ Ori C as imaged with MagAO's Visible CCD camera (VisAO) in different filters. Note the excellent resolution in the raw 60 second image. We note that post-detection shift and add (SAA) was not applied, nor was there any frame selection used to produce these top row images. Typically we achieved resolutions of 0.026-0.029" and Strehls of 28-35% in 0.5-0.7" V-band seeing.
*Middle row:* same as top row, except the images have been post-detection aligned (SAA) and the pixel response function (PRF) has been removed. This improved image resolution by~5-6 mas.
*Bottom row:* is just magnified by 3x These were the highest resolution, deep, images ever published to our knowledge (modified from Close et al. 2013).

A simple reason that many "VisAO" science cases *cannot* be done with *HST* is that the brightest science targets (V<8th mag) become difficult to observe in space without debilitating core saturation/charge bleeding --even for minimum exposures. Moreover, the permanent loss of the ACS HRC channel leaves just the visible coronagraphic "wedge" in STIS on *HST*. With a size of >0.2", which often covers up the most important science area for circumstellar science (the core), this bar inhibits HST study of our "VisAO" science cases. Also we note that *JWST* will not likely produce diffraction-limited imaging in the visible. MagAO with its VisAO camera provides a 2-3 fold improvement in the angular resolution of direct imaging in astronomy while simultaneously gaining access to the important narrowband visible (0.6-1.05 μm) spectroscopic features (like the key hydrogen recombination Hα line at 0.656 μm) that have been inaccessible at 0.02" resolutions to date (Close et al. 2013; Close et al. 2014). Similar resolutions are now also possible with SPHERE/ZIMPOL (see Fusco et al. 2016).

## 1.5 Robotic AO Systems and Ground Layer AO (GLAO) in the Visible

Visible AO is being used on smaller telescopes still today. An excellent example of this is the Robo-AO system in use at the 1.5m at Palomar. It uses an UV LGS to allow remote scanning of the skies. Robo-AO can do thousands of AO targets in a survey at an unprecedented (for AO) speed of ~15 targets/hour. Although the spatial resolutions are limited ~0.2" the ability to image thousands of targets in a survey is very powerful. For more on Robo-AO see Baranec et al. 2016 and for more on its new home at the 2.0m on Kitt Peak see Salama et al. 2016 (these proceedings). Robo-AO completed a follow-up survey of 3313 of the of *Kepler* field KOI's (and the discovery of 440 companions) as an excellent example of visible AO's scientific power (in fig 2: Law et al. (2014) detected 53 companions of 715 KOI targets; Baranec, et al. 2016; 203 companions from 969 KOI; and Ziegler, et al. 2016; 223 companions from 1629 KOIs).

Actually host star characterization with AO in visible was first started at the 4m Advanced Electro-Optical System telescope (a rare example of visible AO science pre-2013), in Maui with the high-order AO system there, see for example Roberts et al. (2011; 2015). However, Robo-AO can cover more stars faster than any other system, and currently dominates the host star follow-up field.

Visible GLAO can be a difficult niche to exploit since GLAO images are worse than what HST can deliver, yet HST time is very limited, and visible GLAO can be advantageous over seeing limited observations. Excitingly, recent first light measurements by ESO's AOF (Oberti et al. 2018 these proc.) show a factor of 2x FWHM improvement over the 60x60" FOV with MUSE. Moreover, 'imaka at first light (Chun et al. 2018; these proc. and Lu et al. 2018; these proc.) show similar levels of correction over VERY large 11x11' FOV. Also Lu et al. (2018; these proc.) show hints that GLAO improvement can be significant all the way to the B band!

There are other visible AO instruments on even larger telescopes like those that are fed by the very high order PALM-3000 AO system at the 5m Hale at Palomar. Examples of its visible science cameras include the SWIFT IFU and TMAS CCD camera which has made excellent images of the moons of Jupiter (Dekany et al. 2013).

## 1.6 The Largest Aperture Visible AO System: The SPHERE/ZIMPOL Polarimetric Camera

The largest telescope visible AO system is the ZIMPOL camera fed by the SPHERE ExAO system (~1.2Khz, ~1300 sub-aperture SH WFS; Fusco et al. these proceedings). In good conditions SPHERE can achieve moderate Strehls well in the visible. For example this has allowed Kervella et al. (2016) to do fast integrations over 4 different visible wavelengths (V, CntHα, NHα and TiO717) with SPHERE/ZIMPOL and resolve the stellar surface of Betelgeuse. They detected an asymmetric gaseous envelope inside a radius of 2 to 3 times the near infrared photospheric radius of the star ($R_*$), and a significant Hα emission mostly contained within 3 $R_*$. From the polarimetric signal, they also identified the signature of dust scattering in an asymmetric and incomplete dust shell located at a similar radius. The presence of dust so close to the star may have a significant impact on the wind acceleration through radiative pressure on the grains. The 3 $R_*$ radius emerges as a major interface between the hot gaseous and dusty envelopes. The detected asymmetries strengthen previous indications that the mass loss of Betelgeuse is likely tied to the vigorous convective motions in its atmosphere.

Similar work has now also been done on the AGB star envelopes around R Dor (Khouri et al. 2016) and L2 Puppis (Kervella et al. 2014).

There is also the powerful use of combining the fact that dust scatters light strongly at lightly less than ~1μm (similar to the <1μm size of the dust particles according to Mie scattering theory). In addition, ZIMPOL utilizes a differential approach to the detection of the polarized light that allows for the rejection of all non-polarized light.

Moreover, polarized light data reduction itself can also reject the non-centrosymmetrically polarized light – to reveal only the proper scattered centrosymmetric polarized ($Q_\varphi$) light from a circumstellar disk (see Fig. 2). There are also more visible AO cameras coming on-line soon.

Subaru's SCExAO (Jovanovic et al. 2016) has already been producing high Strehl science in the NIR and will soon be producing science in the visible with VAMPIRES (Norris et al. 2016) and FIRST (see Jovanovic et al. 2016) with its 3.5kHz control of a woofer (AO188 DM) and tweeter (BMC 2Kilo DM) ExAO system.

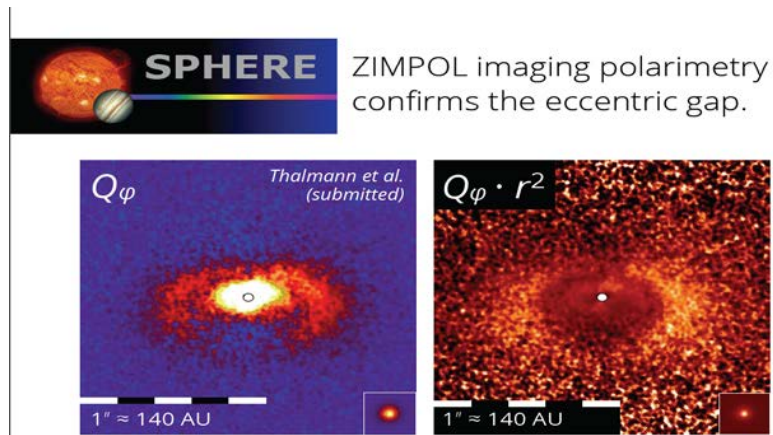

*Fig. 2:* Here the scattered light around the T-Tauri Star LkCa 15b as traced by the centrosymmetric polarized light detected ZIMPOL in the visible (Thalmann et al. 2015). Note the disk PA is really at around 230$^o$, compare to Fig. 12.

## 2.0 MagAO

In this section we introduce in more detail the first large telescope AO system (MagAO) to produce visible AO science during most of its on-sky time. To best understand how the MagAO system works in the visible first need to understand a bit about the history of the Adaptive Secondary Mirror (ASM).

### 2.1 Past Developments of Adaptive Secondary Mirrors for Adaptive Optics

Adaptive secondary mirrors (ASMs) have several advantages over conventional deformable mirrors: 1) they add no extra warm optical surfaces to the telescope (so throughput and emissivity are optimal); 2) the large size of the optic allows for a relatively large number of actuators and a large stroke; 3) their large size enables a wide (>5') field of view (FOV); 4) the non-contact voice-coil actuator eliminates DM print-through; and hence 5) performance loss is minor even if up to ~10% of the actuators are disabled (proof: all MagAO data in this paper was obtained with ~12 of 585 MagAO actuators disabled). They also give better "on-sky" correction than any other AO DM (see Fig. 1). Hence, adaptive secondaries are a transformational AO technology that can lead to powerful new science and telescopic advancement (Lloyd-Hart 2000). MagAO is the result of 20 years of development by Steward Observatory and our research partner INAF-Osservatorio Astrofisico di Arcetri of Italy and industrial partners Microgate and ADS of Italy.

In 2002 this Arizona/Italy partnership (Wildi et al. 2002) equipped the 6.5m MMT with the world's first ASM. This ASM is a 65 cm aspheric convex hyperboloid Zerodur shell 2.0 mm thick. The thin shell has 336 magnets bonded to its backside where 336 voice-coil actuators with capacitive sensors can set the shell position. The MMTAO has carried out regular NIR science observations since 2003 reliably with little down time (see for example: Close et al. 2003a; Kenworthy et al. 2004, 2007, 2009, Heinze et al. 2010; Hinz et al. 2010). However, the MMT system was really a prototype ASM.

From the many lessons learned from the MMT's ASM, a new "2$^{nd}$ generation" of ASMs was fabricated for the LBT and MagAO. LBT's AO system has had a spectacular on-sky first light in June 2010 (Esposito et al. 2010) obtaining the best AO performance of a large telescope to date. ESO has also developed (a larger 1.1m, thicker ~2.0 mm) DSM shell for their future AO facility for science use in late ~2016 (AOF; see Arsenault et al. 2014;

Madec et al. 2016; these proceedings). Also such ASMs are baselined for the 24m Giant Magellan Telescope (GMT) secondaries (~2024) and the M4 of the ~39m E-ELT (~2026), and perhaps as a future upgrade to the secondary of the 30m TMT. *Adaptive secondaries are now key to major AO systems and will likely play a role in all future large telescope projects.*

## 2.2 The 2$^{nd}$ Generation 585 element ASM for MagAO

Our "thin shell/voice coil/capacitive sensor" architecture is the only proven ASM approach. MagAO's "LBT-style" 2$^{nd}$ generation 585 actuator 85cm dia. ASM offers many improvements over the 1$^{st}$ generation "MMT" ASM. In particular, MagAO's successful Electro Mechanical acceptance tests in June 2010 proved that the MagAO 585 ASM has larger stroke ($\pm15$ µm), a thinner shell (at 1.6 mm vs. 2.0 mm), half the "go to" time (<0.7-1.0 ms; with electronic damping), 2-5 nm rms of positional accuracy (by use of a 70 kHz capacitive closed-loop), and just 30 nm rms of residual optical static polishing errors (compared to ~100 nm rms on the current MMT shell). These improvements are taken advantage of by LBTAO as well (Riccardi et al. 2010), but MagAO's mirror overall is slightly better behaved compared to LBT (MagAO's lack of 87 slower "outer ring" LBT actuators increases its speed w.r.t LBT). Moreover, MagAO's ASM is much more flexible than any other ASM, while also not having the inner "stressed" hole illuminated (due to the 0.29 central obscuration of Magellan). So it is not really surprising that MagAO should be the highest performance ASM yet built – see Fig. 1 for proof of how effective MagAO is at high-order correction.

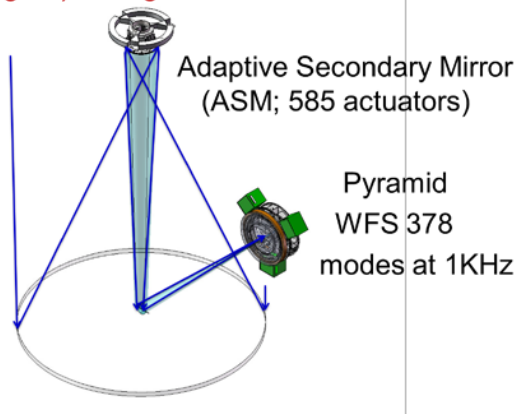
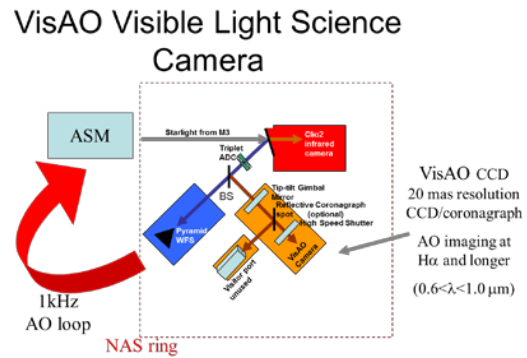

*Fig 3:* Schematic diagrams of the MagAO systems and VisAO camera.

Magellan is a Gregorian telescope (Fig. 3 -*left*) and requires a large (d=85.1 cm) concave ellipsoidal ASM. The concave shape of a Gregorian secondary enables easy testing "off the sky" with an artificial "star" for daytime tests and interferometric recalibration in the closed telescope dome. In addition, 585 mode servo loop CPU latency is limited to <120 µs through the use of 132 dedicated DSPs (producing ~250 Giga Flops, in the ASM electronics) for very fast real-time performance.

## 2.3 Current Status of the MagAO System

MagAO had first light (with its science cameras VisAO and Clio2; see Fig. 3-*right*) in November 2012 (as scheduled after the PDR that occurred in 2009). During first light the AO performance was excellent in the visible with just 200-250 modes in closed loop (see Close et al. 2013). The system than had a major upgrade in the second commissioning run in April 2013 when the maximum number of corrected modes was raised to 378 and corrections as good as 102 nm rms were obtained on bright stars in median 0.6" seeing. Then from April 1 to 25$^{th}$ (in the 2014A semester of Magellan) the first open facility science run was executed. The demand for MagAO observing time in 2014A was very high (typically 3-5x oversubscribed where it has remained). The individual TACs from Carnegie, MIT, Harvard/CfA, Arizona, University of Michigan, and Australia all assigned time for their users on MagAO in 2014A. The 2014A run was a success with almost all science programs executed and with less than 2% downtime from MagAO and its cameras. Similar results from 2014B, 2015A, 2015B campaigns. Overall the community was very happy with MagAO's performance. To date these TACs have now assigned a total of ~150 nights to MagAO.

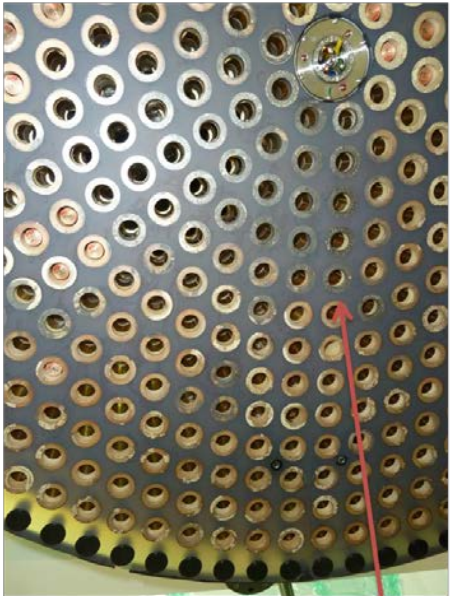
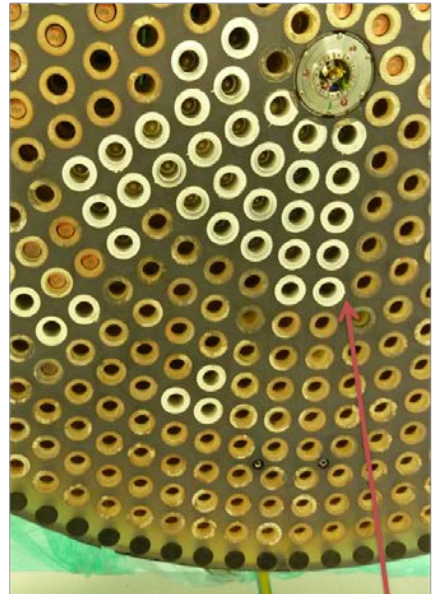

**Fig. 4a:** The reference body before and after silver coating repair for capacitive sensors.

During the 2016A run, on Feb 19 2016, a small ~2 liter glycol leak in the ASM contaminated the ASM. It has since been cleaned and fully restored and tested. The total number of bad positional sensors after the leak was 47 (all in the same area). We were able to develop a custom technique of applying "spray silver" coating to the reference body (https://www.youtube.com/watch?v=zeBMgkH7sj4) to replace the silver (really silver sulfide, $Ag_2S$) that was lost around 41 of these actuators (Fig 4a). The coatings were lost due the electrolytic reaction where the $Ag_2S$ is removed once the liquid alkaline glycol connects circuit between the $Ag_2S$ and Al (coated on the shell backside). Below are the details:

**Due to the classic Electrochemistry, oxidation/reduction that occurs during a Glycol leak in ASM gap**

OXIDATION:

$2 \mathbf{Al}(s) + 6 OH^- (aq) \longrightarrow \mathbf{Al_2O3}(s) + 3 H_2O (l) + 6 e^-$

REDUCTION: $\mathbf{Ag_2S}(s) + 2 H_2O (l) + 2 e^- \longrightarrow 2 \mathbf{Ag}(s) + H_2S (aq) + 2 OH^- (aq)$

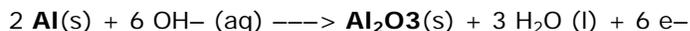

$3 \mathbf{Ag_2S}(s) + 2 \mathbf{Al}(s) + 3 H_2O (l) \longrightarrow 6 \mathbf{Ag}(s) + 3 H_2S (aq) + Al_2O_3(s)$

This reaction strips the silver sulfide $Ag_2S$ (which is really the entire armature since it is nearly 100% $Ag_2S$ after ten years when less than 1micron thick). If the reaction is very fast and aggressive (with the 10 V ASM power on – shell "set") the individual "fresh" Ag particles are lost to the solution (or will form a dark stain from the fresh silver deposit – which turns rapidly back to $Ag_2S$ in the water once the 10V is off) on the backside of the shell) as it drains out. This new layer of $Ag_2S$ can be easily wiped off the back of the shell and also leaves a dark powder on the reference body holes and gap surface.

Also the Sulfur will leave the Silver for the Aluminum to produce white $Al_2O_3$ solid on the back of the shell, but also leave some parts of the backside of the shell with no Al coating at all. This reaction will only go well in the high pH of the glycol (MagAO's glycol has a pH of >8), hence in pure $H_2O$ there is lower pH and the reaction stops removing silver quickly. Moreover, if the 10 Volts is applied across gap as the glycol is introduced this will drive the $Ag_2S$ over to the Al even faster (as was the case with the Feb 19, 2016 Glycol leak with MagAO).

A previously scheduled (MagAO-2K, PI: Jared Males) 2016B interferometric re-calibration of the ASM occurred, as scheduled, on Nov. 2016 (fig 4b) see Briguglio et al. 2018 for details. After which MagAO was fully operational again, see Fig *4b – bottom* of a 25 mas FWHM PSF at 0.64µm). Normal science runs have now occurred in 2017A and 2018A.

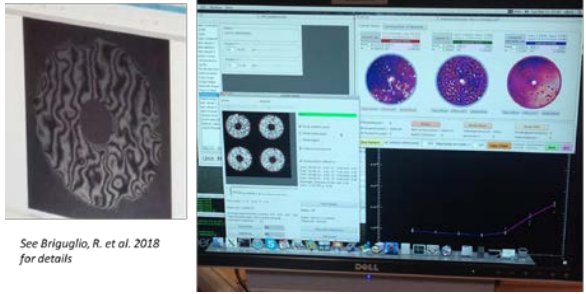

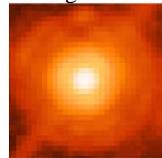

*Fig 4b:* The interferometric recalibration of the ASM (left). Also note the new 30 pixel diameter Pupils (all 4 of them) of the newly upgraded PWFS from the MagAO-2K upgrade (Males et al 2016). The new, post-repair, IMATs allow excellent closed-loop performance in the visible ~100-130 nm rms on bright NGS in median conditions. Post–repair PSF (FWHM 25 mas at 0.64 µm) Wu et al. 2017b.

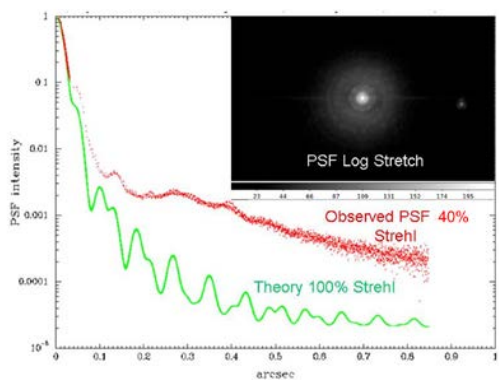

*Fig 5:* Deep PSF (40% Strehl) at 0.98 µm PSF (140 nm rms WFE with just 200 modes). Modified from Close et al. 2013. Note how the 35 mas resolution of the PSF allows for a raw contrast of 500 at just 0.1" separation. In this manner good inner working angles and contrasts can be achieved with visible AO even if the Strehls are lower than in the NIR.

The schematic drawing (Fig. 3 –*right-hand side*) outlines how our VisAO and Clio2 cameras are co-mounted and can be used simultaneously (if desired). No instrument changes are ever needed to switch between NIR (Clio2) and visible science (VisAO). In the campaign/queue mode used for MagAO one will not use VisAO if seeing is poor, nor will any telescope time be lost since 1-5.3 µm Clio2 science can be done in >90% of the seeing conditions at Las Campanas --an excellent (median V=0.64" ) seeing site (Thomas-Osip 2008). See Morzinski et al. 2016 for more details about MagAO and in particular for details about its IR (1-5μm) camera Clio.

## 2.4 Our Simulated MagAO/VisAO Error Budget Compared to On-Sky results

MagAO's 585 controllable modes map to a 22 cm "pitch" on the 6.5m primary. To predict the exact degree of correction we used "end-to-end" simulation of MagAO/VisAO with the Code for Adaptive Optics Simulation (CAOS; Carbillet et al. 2005). Our CAOS simulations (assuming no extra telescope vibration) predict slightly larger wavefront errors (135 vs. 122 nm rms) than our typical lab results (Males et al. 2012; Close et al. 2012).

### *2.4.1 Moderate Strehls robustly obtained in the Visible with MagAO on-sky*

While simulations, and lab tests give significant comfort, the true test is whether the AO system can actually achieve >30% Strehls reliably in median atmospheric conditions (0.7") on the sky. Since fall 2012 MagAO has been used over 150 nights. Our published results show that the Strehls achieved by MagAO/VisAO are right along the expected values. We now have >10 MagAO runs with diffraction-limited images obtained with VisAO on most of the nights. Diffraction-limited visible imaging with MagAO/VisAO is almost guaranteed at Magellan in median seeing/wind conditions on V<11 mag guide stars (Close et al. 2014). To quantify this, MagAO has achieved 104-140 nm of wavefront error using 300 modes in 0.65" seeing (median conditions) on sky. This performance is very similar to the predictions by Males et al. 2013 for MagAO.

## 2.5 A Few Selected Science Cases for Visible AO Imaging (MagAO/VisAO) Observations

### *2.5.1 Imaging of Young Stars and Proplyds in Orion*

Clearly the exciting possibility of obtaining ∼20 mas FWHM images with MagAO could enhance our understanding of the positions (and motions) of the nearest massive young stars. Hence we targeted the Orion Trapezium cluster during the first light commissioning run with the MagAO system. Indeed in Figure 1 we can see how well the 32 mas binary Theta 1 Ori is split with visible AO (Close et al. 2013). As we can see in Fig. 6 in the trapezium region alone many scientific results can be obtained with visible AO. We can measure the astrometry of the trapezium members down to 0.2 mas/yr --which is excellent (Close et al. 2013). This translates to velocities of 0.4 km/s, at which point the individual orbits start to show orbital arcs in the Trapezium (Close et al. 2013).

As Fig. 6 also illustrates we can image off-axis proplyds at high resolution. These warped protoplanetary disks are being photoevaporated by the intense UV radiation and solar wind from theta 1 Ori C. Interestingly Wu et al. 2013 were able to use it as a guide star to make HST resolution images of LV1 6.3" off-axis from the guide star. This suggests (as do the ~10" corrected FOV from SCAO solar AO systems) that the corrected FOV in the optical can be (even at Halpha) as large as 6" (Wu et al. 2017c).

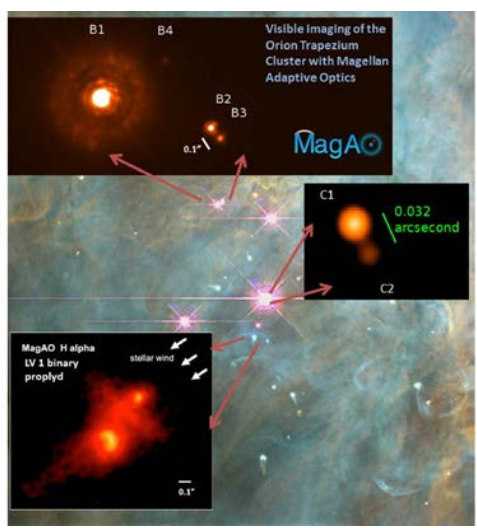
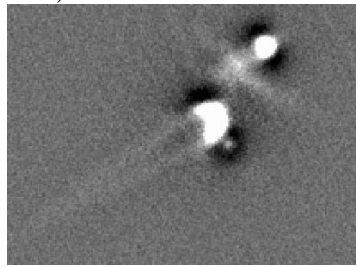

*Fig 6:* High resolution images of the young stars (and their orbits) in the famous Trapezium cluster in the Orion Nebula. Close et al. 2013 and Wu et al. 2013 (background image from NASA/*HST* archive). Below we show a more recent image of LV1 from Wu et al. 2017c (note the Ha "point source" at PA ~200).

*2.5.2 High-Contrast Imaging with VisAO*

Another Key science field for visible AO is high contrast imaging of exoplanets with direct imaging. The large ~10 Jupiter mass planet Beta Pic b some ~9 AU from its host star is an excellent target for this sort of work. At the first light of MagAO deep imaging of Beta Pic b was carried out to see how faint the planet would appear in the Ys band. The full results of this first detection of an exoplanet with a CCD from the ground can be found in Males et al. (2014). In Fig. 7 we show what the contrast curve was from 2.5 hours of open shutter observation of the system. We note that this achieved contrast is very similar to that obtained by GPI on Beta-Pic in 30min in H band with SSDI/TLOCI (http://www.gemini.edu/sciops/instruments/gpi/instrument-performance?q=node/11552 ). The important fact here is that the smaller $\lambda/d$ of visible AO can compensate for poorer (40%) Strehls to enable similar contrasts of an "extreme AO" system like GPI working at H band.

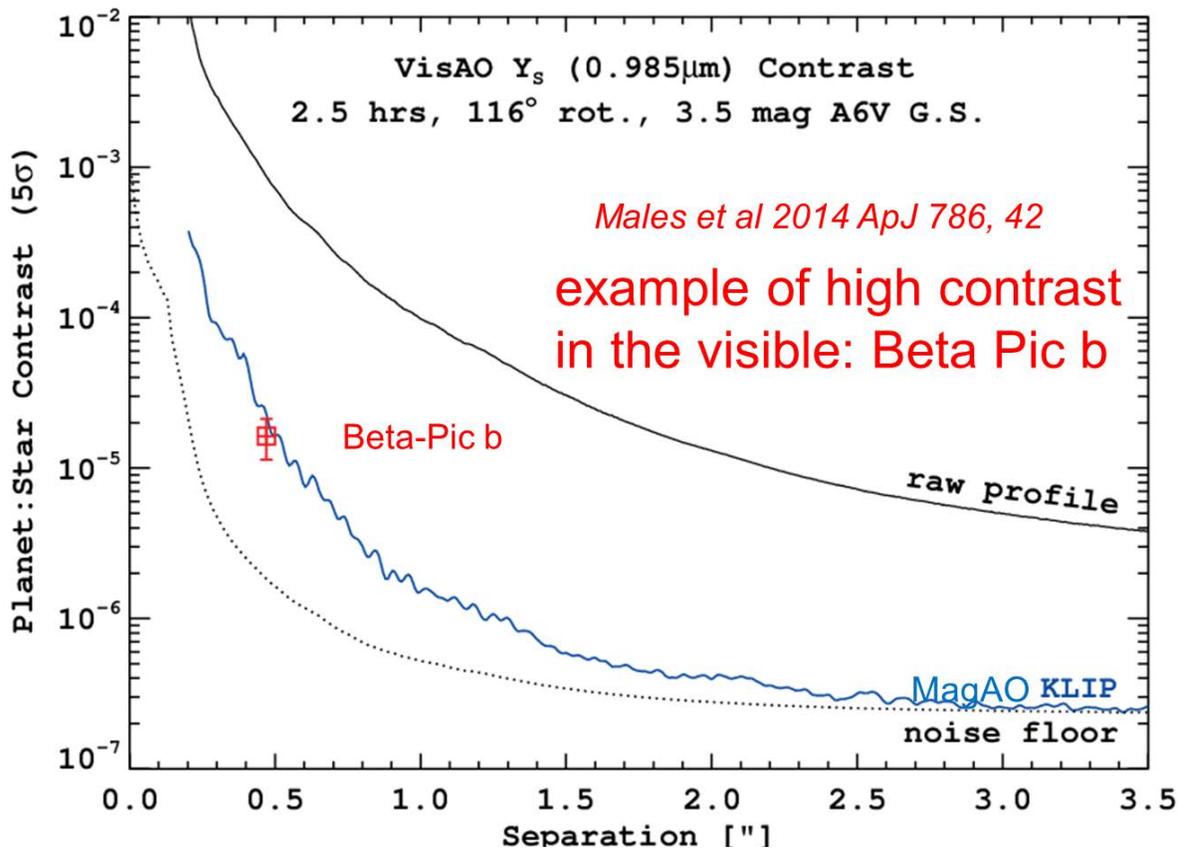

*Fig 7:* This "first light" visible AO contrast plot for MagAO from the published results of Males et al. (2014). It shows a similar contrast as that of the Gemini Planet Imager (GPI) at H band in 30 min (also on Beta-Pic). Visible AO is also "extreme AO" in terms of contrast.

*2.5.3 High Contrast imaging with a novel SDI Camera*

In the last section we learned that visible AO can make high contrast images in the visible broad band with ADI/PCA reductions. Indeed the VisAO camera design incorporates all the key features and remotely-selectable elements necessary to optimize such visible AO science. In particular, the coronagraph wheel contains a range of our custom reflective ND masks (Park et al. 2007), allowing deep circumstellar science on bright targets that would otherwise saturate the detector. The main purpose of these masks is to prevent blooming of the CCD47 VisAO detector. Our coronagraph doesn't need to suppress diffraction rings since the ADI/PCA reduction technique work very well. However the number of useful astrophysical investigations possible with Strehls in the

~20-40% range are limited without accurate PSF calibration. In the next section we introduce the "SDI" mode of the VisAO camera to calibrate the PSF directly.

One needs simultaneous PSF information to compare to (or subtract against) the "in-line" science image. An extremely effective technique for this is Simultaneous Differential Imaging (SDI; Close et al. 2005) which utilizes a Wollaston Beamsplitter to obtain nearly identical, simultaneous, images of the *o-polarized* and *e-polarized* PSFs (typically there is <10 nm rms of non-common path SDI error between the *o* and *e* images; Lensen et al. 2004; Close et al 2004). The SDI configuration of the VisAO camera includes: 1) a thin small angle calcite Wollaston beamsplitter near our pupil; and 2) a split on-Hα/off Hα "SDI" filter just before the focal plane (for the *e* beam/*o* beam; see Fig. 8). In this mode, we have obtained an almost perfect (photon noise limited) simultaneous calibration of the PSF "off" and "on" the Hα line on-sky. Hence, a simple subtraction of the "off" image from the "on" image will map Hα structures (jets, disks, accreting faint companions etc.), with minimal confusion from the continuum or PSF. We have three such SDI filter sets (with optional SDI double spot coronagraphic masks) for the Hα, [OI], and [SII]. Our new (as of April 2014) Halpha SDI filters yield on-sky 1 hr. images of $2 \times 10^{-6}$ contrast at 1" which exactly at the photon-noise limited floor. However this assumes the source is not polarized itself.

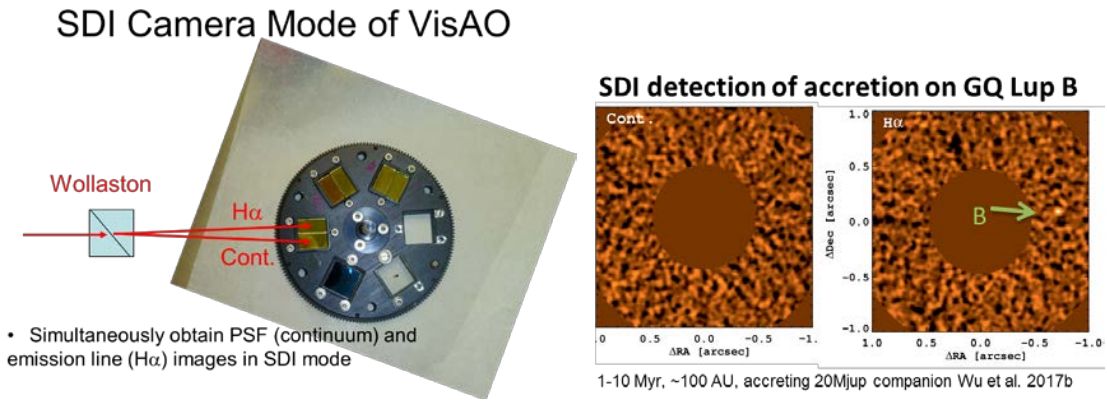

*Fig 8:* A cartoon of the SDI mode (left). To the right we see a real science example of the MagAO SDI mode in action. Here we see how a young forming solar system has its giant planet/brown dwarf detected by light of gas accreting to its surface where it is shocked and glows in Hα. Modified from Wu et al. 2017a.

*2.5.4 High-Contrast Hα Emission Line Imaging of Young Extrasolar Planets: GAPlanetsS Survey*
Another key survey project we are doing with VisAO is high-contrast Hα SDI imaging of the emission from the accretion shock caused by gas accreting onto gas giants during formation. We'll target >30 known southern, young (<10 Myr), nearby (<150pc), gas-rich transitional disk systems that have I≤10 mag. During the time of gas accretion the protoplanets will have ~$10^{-(3-4)}$ $L_{sun}$ and much of this will be radiated at Hα for ~$10^6$ yrs (Fortney et al. 2008). In this manner, we can finally directly image giant planets where/when they form (likely past the "snow-line" >30-50 mas) as Hα point-sources orbiting the young target star. This GAPplanetS survey is already 80% complete, and will be likely finished after the 2019A run this spring (Follette et al. 2018 in prep).

The first major result of GAPlanetS was the detection of the accretion low mass star HD142527B (Close et al. 2014) inside the disk gap of HD142527A – see figure 9. From this image the accretion rate could be determined for the companion that is responsible for carving out a large gap in the disk of HD142527A.

The next science highlight was the first detection of a true protoplanet. In the LkCa 15 system Sallum et al. (2015) detected with MagAO an H-alpha source at 90 mas at the same position of a L' band NRM detected point source from the LBT and Keck (Krause et al. 2010). This discovery proves that planets pass through a high-luminosity accretion phase early in their lives and also have strong magnetic fields then. See Fig. 10 for discovery Hα images.

**2.5.4.1 Problem with SDI:** there is a weak systematic error with SDI when looking at sources that are polarized themselves. Dust clumps can be 1% star brightness. If the clump is 10% polarized. Then you have a clump at $10^{-3}$ contrast which might pass through one arm of the Wollaston and lead to one part of the Ha PSF being 1/1000 brighter than the Continuum PSF is at that field position– that limits contrast targets with dusty disks.

**2.5.4.2. SDI+ Solution:** *Depolarize the input beam.*
We now rapidly spin a half wave plate at 2Hz, which evens out the clump in the 2 Wollaston produced beams by modulating the differential clump brightness at 8Hz (Fig. 11). In a 5 second science exposure we'd average over 40 cycles – so in integrated light the clump SDI signal (Hα-Continuum) is only $1/1000/40 = 2.5 \times 10^{-5}$ per 5s frame. See figure 12 and 13 for examples of the contrasts now possible with SDI+. Thanks to SDI+ and its 2-3x higher contrast (Fig. 13) we were able to recover several (3) new protoplanets in Hα (see Follette et al. 2018; Wagner et al. 2018). SDI+ has allowed us to conclude that accreting gap planets are bright and relatively common.

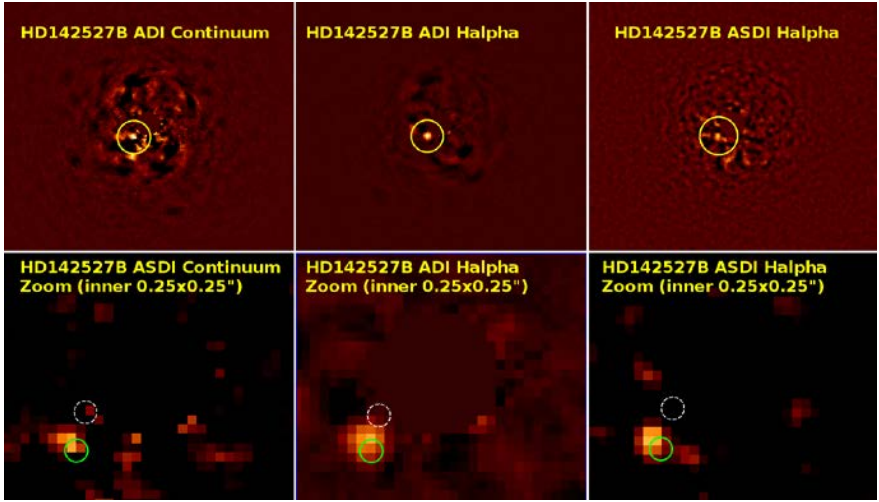

**Fig 9:** Here we examine the first GAPplanetS target, the famous transitional disk HD142527 with the SDI mode of MagAO. (*left*) Continuum (643 nm) ADI reduced image. (*Bottom Zooms*): Note the weak (~3 sigma) detection near the location of the candidate of Biller et al. 2012 (green circle). The source is inconsistent with the background star position (white circle). (*middle*) Hα ADI images. Note the unambiguous 10.5 sigma Hα point-source at sep=86.3 mas, PA=126.6°, hereafter HD142527B. (*right*) ASDI data reduction, here NCP narrow-band filter ghosts are not as well removed as with ADI. NOTE: the new single substrate SDI filters designed by custom scientific (and installed in VisAO in April 2014) has removed all signs of filter ghosts in the current Halpha SDI camera. These images clearly detect an accreting young companion just 86 mas (and 1000x fainter in continuum) from the primary star! Modified from Close et al. (2014).

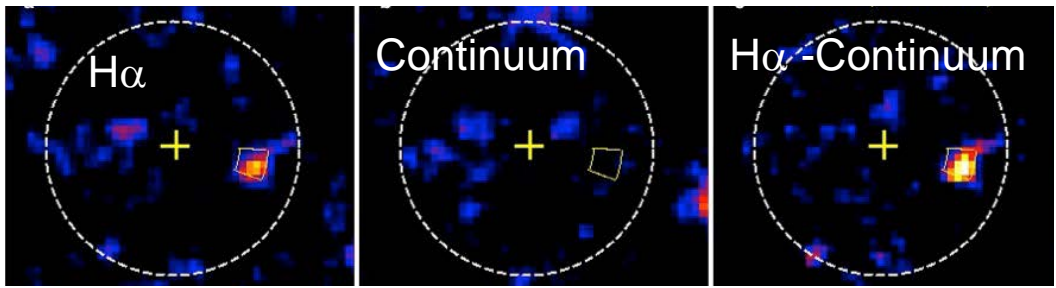

**Fig. 10:** The first detection of a protoplanet. Here we see the MagAO SDI Hα detection of the 2-5 Jupiter mass protoplanet LkCa 15b. The planet is at ~15 AU (90 mas) and was unambiguously determined to be accreting by detection of Hα emission. These images were published in the Nov 19, 2015 issue of Nature by Sallum et al. 2015. This exoplanet is helping carve out the disk gap seen by Thalmann et al. 2015 in figure 4. Image produce by Kate Follette (modified from Sallum et al. 2015).

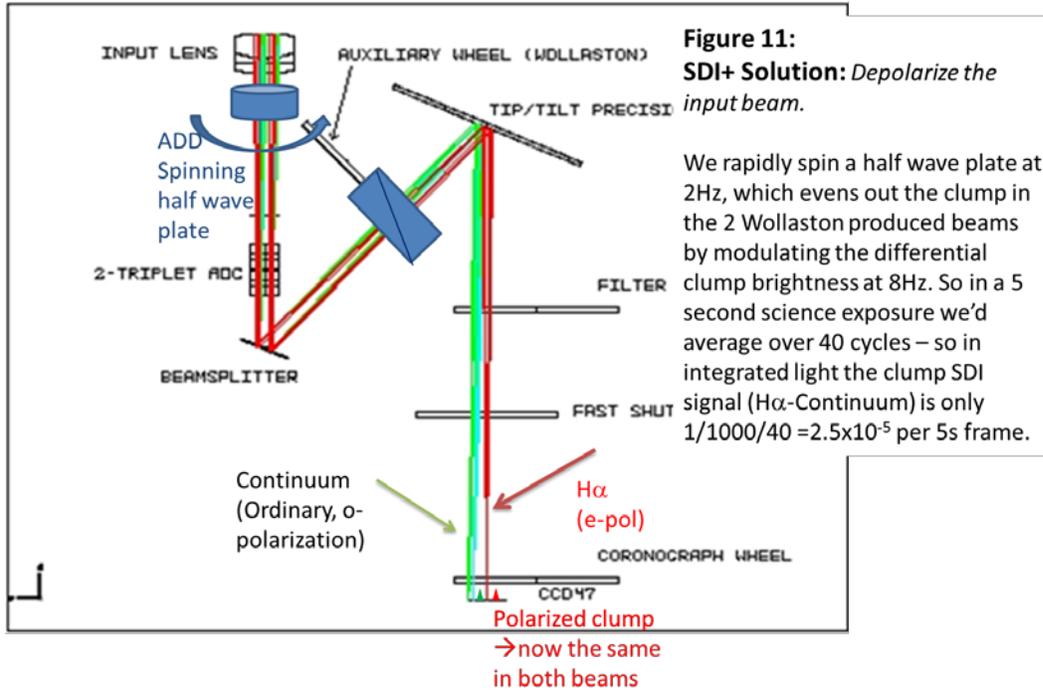

**Figure 11:**
**SDI+ Solution:** *Depolarize the input beam.*

We rapidly spin a half wave plate at 2Hz, which evens out the clump in the 2 Wollaston produced beams by modulating the differential clump brightness at 8Hz. So in a 5 second science exposure we'd average over 40 cycles – so in integrated light the clump SDI signal (Hα-Continuum) is only $1/1000/40 = 2.5 \times 10^{-5}$ per 5s frame.

**Fig 12:** examples of fake planet sensitivity with SDI+

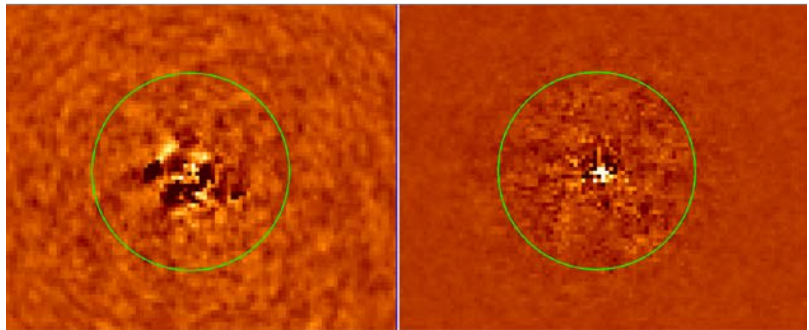

**Fig 13:** Improvement in contrast with SDI+ compared to SDI on the same R~11 dusty star (TW Hya).

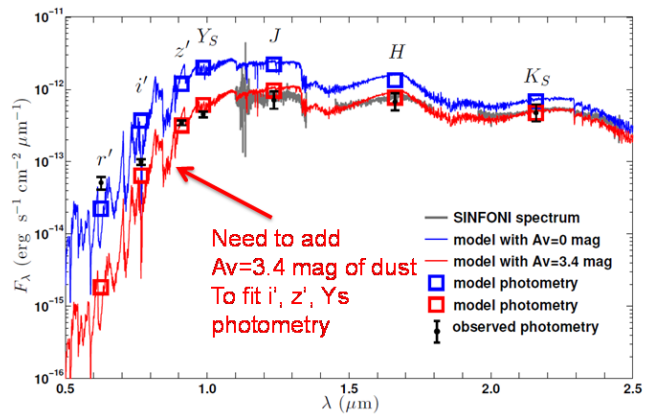

*Fig 14:* Atmospheric model fits to photometry is a powerful tool to understand the complete luminosity of a close brown dwarf companion (Wu et al. 2015). Above we see Wu et al. find with MagAO an Av=3.4+/-1.1 mag of extinction and signs of active accretion, hence CT Cha B has its own disk and is a brown dwarf formed by core fragmentation at 430 AU (figure modified from Wu et al. 2015).

### 2.5.5 Using Visible and NIR fluxes to constrain Atmospheric Models and Extinctions

A great advantage of visible AO is to constrain the atmospheric properties of brown dwarfs and extra-solar planets. Only in the visible can we obtain access to fluxes blueward of the black body peak of the spectrum. Hence the combination of r',I',z' and NIR photometry and spectra can strongly constrain the nature of companion brown dwarfs (Fig. 14) and exoplanets (Fig. 15), see for example Wu et al. 2015a,b, Wu et al. 2016.

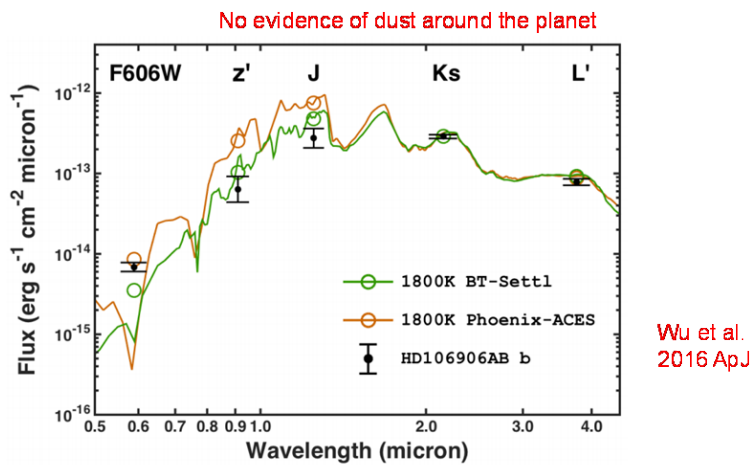

**Fig 15:** Here we see how MagAO visible AO photometry of the exoplanet HD106906b (discovered with MagAO by Bailey et al. 2014) can be proven to be a ~11 Jupiter mass planet through use of evolutionary models fit with the correct extinction of ~0 mag Av (Wu et al. 2016). So unlike the cases of CT Cha B (Wu et al. 2015a) and RX 10609B (Wu et al. 2015b) which were found to be extincted brown dwarfs with their own disks (and not exoplanets), HD106906b on the other hand has no sign of strong extinction from an external dust disk –and so is likely a truly faint, red, exoplanet due to its own cool atmosphere.

## 3.0 FUTURE FOR MagAO

Now that the SDI optics have been upgraded to SDI+ (to maximize H$\alpha$ SDI contrast on polarized objects) no more upgrades to the VisAO camera are planned. Our 2kHz PWFS experiments were successful and will be reported elsewhere (Males et al. in prep.), we have restored MagAO back to just 1 kHz due to read noise issues at 2kHz (30e vs ~8e RON of the CCD39 PWFS).

Recent support from the NSF MRI program is enabling an extreme AO upgrade (MagAO-X) where the MagAO ASM could be simply a "woofer" in concert with a new BMC 2Kilo "tweeter" DM operating at 3.5 kHz. Alternatively, MagAO-X can bypass MagAO completely and use its own internal woofer with Magellan's static f/11 secondary. Visible wavelength very high-contrasts (~$10^{6-7}$) inside 0.5" are the goal of this coronagraphic instrument by 2020. For the MagAO-X system overview please see Males et al. 2018 (these proceedings) and for the optical mechanical design see Close et al. 2018 (these proceedings) for a complete overview of the MagAO-X system.

## 4.0 CONCLUSIONS

The future of visible AO science is clearly very bright. We are now in a new age where AO correction is good enough for excellent visible AO science to be carried out with facility AO systems like Robo-AO, MagAO, and SPHERE/ZIMPOL. With many new visible AO systems on-line right now, like SCExAO at Subaru. There are also wide field visible systems available now: for instance GLAO systems like 'imaka (with a very large 11x11' FOV) and ESO's AOF feeding the 60x60" IFS MUSE ensure many more exciting science results will follow!

We already have exciting MagAO/VisAO science results on *exoplanets atmospheres* (Males et al. 2014; Bailey et al. 2014; Close et al. 2014; Follette et al. 2014; Biller et al. 2014; Morzinski et al. 2015; Sallum et al. 2015;

Rodigas et al. 2016; Wu et al. 2016), *accreting protoplanets* ( Sallum et al. 2015; Follette et al. 2018; Wagner et al. 2018), *young stars* (Close et al. 2013; 2014); *circumstellar disks* (Follette et al. 2013; Wu et al. 2013; Biller et al. 2014; Rodigas et al. 2014; Wu et al. 2016), *brown dwarf atmospheres* (Wu et al. 2015a,b) and *H$\alpha$ emission jets* (Wu et al. 2017c.) with many more science papers on these and other fields soon to be published.). ***Even resolving the H$\alpha$ surface of Stars*** for example the ~40mas surface of Mira A (Sevrinsky et al. 2018 in prep) and the 12mas H$\alpha$ wind envelope of Eta Carinae (Ya-Lin Wu et al. 2017b) have been resolved with MagAO H$\alpha$ SDI. This is just the start of what is sure to be a long-lived productive field of science at visible wavelengths with AO.

L.M.C.'s research with MagAO was supported by the NSF ATI (Grant No. 1506818; PI Laird Close) and is supported by the NSF AAG program #1615408 (PI Laird Close) and NASA's Exoplanets Research Program (XRP; grant NNX16AD44G: PI Katie Morzinski).